\documentclass[aps,prb,reprint,amssymb,amsmath,superscriptaddress]{revtex4-1}

\usepackage{graphicx}
\usepackage[english]{babel}

\begin{document}

\title{Hole transport and valence band dispersion law in a HgTe quantum well with normal energy spectrum}

\author{G.~M.~Minkov}
\affiliation{Institute of Metal Physics RAS, 620990 Ekaterinburg,
Russia}

\affiliation{Institute of Natural Sciences, Ural Federal University,
620000 Ekaterinburg, Russia}

\author{A.~V.~Germanenko}

\author{O.~E.~Rut}
\affiliation{Institute of Natural Sciences, Ural Federal University,
620000 Ekaterinburg, Russia}

\author{A.~A.~Sherstobitov}
\affiliation{Institute of Metal Physics RAS, 620990 Ekaterinburg,
Russia}

\affiliation{Institute of Natural Sciences, Ural Federal University,
620000 Ekaterinburg, Russia}

\author{S.~A.~Dvoretski}

\affiliation{Institute of Semiconductor Physics RAS, 630090
Novosibirsk, Russia}

\author{N.~N.~Mikhailov}

\affiliation{Institute of Semiconductor Physics RAS, 630090
Novosibirsk, Russia}

\date{\today}

\date{\today}

\begin{abstract}

The results of an experimental study of the energy spectrum of the
valence band in a HgTe quantum well of width $d<6.3$~nm with normal
spectrum in the presence of a strong spin-orbit splitting are reported.
The analysis of the temperature, magnetic field and gate voltage
dependences of the Shubnikov-de Haas oscillations allows us to restore
the energy spectrum of the two valence band branches, which are split
by the spin-orbit interaction. The comparison with the theoretical
calculation shows that a six-band \emph{kP} theory well describes all
the experimental data in the vicinity of the top of the valence band.

\end{abstract}


\maketitle

\section{Introduction}
\label{sec:intr}

Peculiarities  of the  energy spectrum of the spatially confined
gapless semiconductors (HgTe, HgSe) results in unique  transport,
optical and other properties of the carriers in the structures with
quantum wells based on such type of materials. Theoretically, the
energy spectrum of confined gapless semiconductors has been intensively
studied since 1981.\cite{Dyak82e,Lin85,Kisin88,Gerchikov90} It was
shown that the linear in quasimomentum ($k$) spectrum (the Dirac-like
spectrum) should be realized at some critical width of the HgTe-quantum
well $d=d_c\simeq 6.3$~nm in the CdTe/HgTe/CdTe
heterostructures.\cite{Bernevig06}  Just these structures attract
especial attention both of theoreticians and of experimentalists.

When the quantum well width $d$ is not equal to $d_c$, the energy
spectrum is more complicated. The calculations show that the
quasimomentum dependence of the carrier energy $E(k)$ for the
conduction band is simple enough both for the wide ($d>d_c$) quantum
wells with the inverse subband ordering and for the narrow ($d<d_c$)
wells with the normal spectrum. The dispersion  is analogous to the
spectrum of the conduction band of usual narrow-gap semiconductors. It
is close to parabolic in the shape  at small quasimomentum (this area
shrinks to zero when the well width tends to $d_c$), crosses over to
linear with $k$ increasing and becomes again parabolic with the further
increase of $k$. The spectrum of the valence band is much more
complicated. Even at $d$ close to $d_c$ it is similar to the spectrum
of the conduction band only within a narrow range of energy near the
top of the valence band. In the wider quantum wells, $d>d_c$, the
spectra of valence and conduction bands differ drastically. At
$d>(10-12)$~nm, the valence band dispersion becomes non-monotonic,
namely, the electron-like section appears at $k\simeq 0$. In narrower
quantum wells, $d<d_c$, the dispersion of the valence band is
nontrivial as well. Together with the main maximum at $k=0$, secondary
maxima in $E(k)$ dependence arise at large $k$ as the theory predicts.

The experimental study of the magnetotransport, energy spectrum  and
their  dependence on the well width  became possible only 10-15 years
ago.  This is primarily due to the impressive progress in  technology.
\cite{Goschenhofer98,Mikhailov06} Although the experimental studies are
mainly focused  on the investigation of the quantum  and spin Hall
effects,\cite{Bernevig06,Koenig07,Bruene10,Gusev10,Gusev13} there is a
number of
papers\cite{Pfeuffer98,Landwehr00,Ortner02,Kvon11,Minkov13,Zhang02,Zhang04,Gusev11,Yakunin12}
where the spectrum of carriers is studied. However, only four of them,
Refs.~\onlinecite{Landwehr00,Ortner02,Kvon11,Minkov13}, are devoted to
the valence band.  One of the key results of the papers is that the
dependence $E(k)$ in the structures with $d>(10-15)$~nm is really
non-monotonic and the electron-like section really appears at $k\simeq
0$. This result is in a qualitative agreement with the calculated ones,
however there are very significant quantitative discrepancies with
theoretical predictions.\cite{Minkov13}

The energy spectrum of the valence band in the structures with the
normal spectrum ($d<d_c$) was experimentally studied   only in
Ref.~\onlinecite{Ortner02}. The measurements were performed at fixed
hole density $p=3\times10^{11}$~cm$^{-2}$, therefore the interpretation
of the data does not seem very reliable.

In this paper, we report the results of  experimental study of the hole
transport in the HgTe quantum well with the normal energy spectrum. The
measurements were performed over a wide range of  hole densities.
Analysis of experimental data allows us to reconstruct the energy
spectrum of the $H1$ hole subband, which has been shown to be in good
agreement with the results of the \emph{kP} theory.

\section{Experimental}
\label{sec:expdet}

Our HgTe quantum wells  were realized on the basis of
HgTe/Hg$_{1-x}$Cd$_{x}$Te ($x=0.55-0.65$) heterostructure grown by
molecular beam epitaxy on GaAs substrate with the (013) surface
orientation.\cite{Mikhailov06} The sketch of the structures
investigated is shown in the inset of Fig.~\ref{f1}(a). The nominal
width of the quantum well was   $d=5.8$~nm,  $5.6$~nm, and  $5.0$~nm,
in the structures H724, H1122 and H1310, respectively. The results for
all the structures  are similar and we will discuss the results which
were obtained in the structure H724 with the higher Hall mobility. The
samples were mesa etched into standard Hall bars of $0.5$~mm  width and
the distance between the potential probes of $0.5$~mm. To change and
control the hole density $p$ in the quantum well, the field-effect
transistors were fabricated with parylene as an insulator and aluminium
as a gate electrode. The measurements were performed at temperature of
$1.3-4.2$~K in magnetic field $B$ up to $7$~T.

\section{Experimental results and discussion}
\label{sec:res}

\begin{figure}
\includegraphics[width=1.0\linewidth,clip=true]{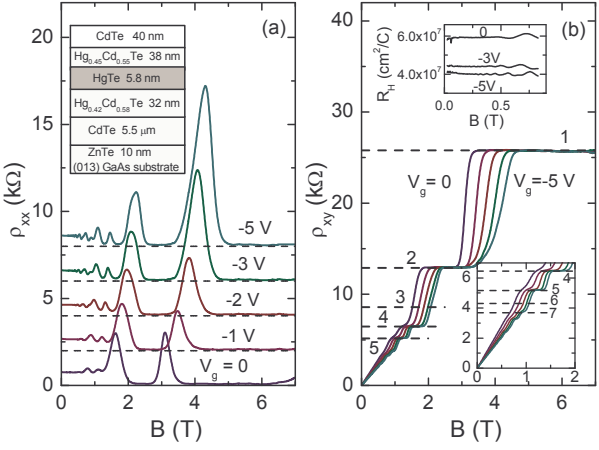}
\caption{(Color online) The magnetic field dependences of  $\rho_{xx}$ (a) and  $\rho_{xy}$ (b)
measured for the different gate voltages. The inset in (a) shows the sketch of the structure H724.
The upper inset in (b) demonstrates the magnetic field dependence of the Hall
coefficient at low magnetic field; the lower inset shows  $\rho_{xy}$ at $B<2$ T}\label{f1}
\end{figure}

\begin{figure}
\includegraphics[width=0.78\linewidth,clip=true]{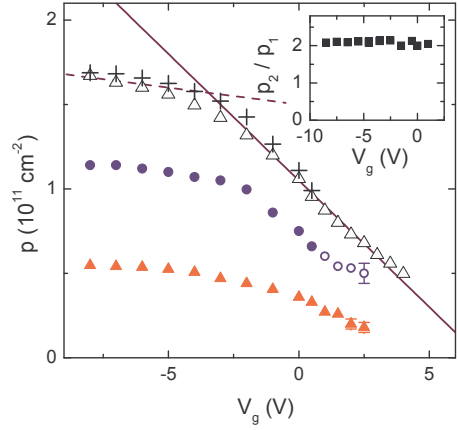}
\caption{(Color online) The gate voltage dependence of the Hall density
$p_{\,\text{H}}=1/eR_{\,\text{H}}(0.1\,\text{T})$ (triangles),
densities $p_1$ and $p_2$ (filled circles and triangles,  respectively) found from the SdH oscillations (see text).
The open circles are $p_2$ found as $p_{\,\text{H}}-p_1$. The crosses are the sum
$p_1+p_2=p_{tot}$. The solid and dashed straight lines are drawn with the slopes
$-1.5\times 10^{10}$~cm$^{-2}$V$^{-1}$ and $-0.2\times 10^{10}$~cm$^{-2}$V$^{-1}$, respectively.
The inset shows the $V_g$ dependence of the ratio $p_2/p_1$.
}\label{f2}
\end{figure}

An overview of the magnetic field dependences of a longitudinal and
transverse  resistivity ($\rho_{xx}$ and $\rho_{xy}$, respectively) for
different gate voltages ($V_g$) is presented in Fig.~\ref{f1}. Well
defined quantum Hall plateaus in $\rho_{xy}$  and minima in $\rho_{xx}$
are evident. It should be noted that the plateaus with the numbers $3$
and $6$ [see lower inset in Fig.~\ref{f1}(b)] are not observed. This
point will be discussed later. As clearly seen from the upper inset in
Fig.~\ref{f1}(b), the Hall coefficient $R_{\,\text{H}}=\rho_{xy}/B$ is
practically independent of the magnetic field in the low-field domain
$B\simeq (0.01-0.2)$~T, where  the Shubnikov-de Haas (SdH) oscillations
are not observed yet. One can therefore assume that the density of
holes can be obtained as
$p_{\,\text{H}}=1/eR_{\,\text{H}}(0.1\text{~T})$. The gate voltage
dependence of so obtained hole Hall density is presented   in
Fig.~\ref{f2} by triangles. One can see that $p_{\,\text{H}}$ linearly
changes with  $V_g$ with the slope $|dp_{\,\text{H}}/ dV_g|$ of about
$1.5\times 10^{10}$~cm$^{-2}$V$^{-1}$ at $-3.5\text{ V}<V_g<4$~V, where
the hole density is less than $1.5\times 10^{11}$~cm$^{-2}$. At
$V_g\lesssim -3.5$~V, the slope becomes much less;
$|dp_{\,\text{H}}/dV_g|\simeq 0.2\times 10^{10}$cm$^{-2}$V$^{-1}$. Note
the capacitance $C$ between the gate electrode and two-dimensional
channel in this structure is constant over the whole gate voltage range
so that the value of $C/e=(1.4\pm0.15)\times10^{10}$~cm$^{-2}$~V$^{-1}$
is practically the same as $|dp_{\,\text{H}}/ dV_g|=1.5\times
10^{10}$~cm$^{-2}$V$^{-1}$ observed at $-3.5\text{ V}<V_g<4$~V.
Possible reasons for the $|dp_{\,\text{H}}/dV_g|$ decrease evident at
$V_g<-3$~V are considered in the end of this section. The Hall mobility
$\mu_{\,\text{H}}=\sigma R_{\,\text{H}}(0.1\text{ T})$, where $\sigma$
stands for the conductivity at $B=0$, increases with the
$p_{\,\text{H}}$ increase, achieves the maximal value of about $8\times
10^4$~cm$^2$/V~s at $p_{\,\text{H}}=1.3\times10^{11}$~cm$^{-2}$, and
demonstrates slight decrease with the further $p_{\,\text{H}}$ increase
(not shown).

\begin{figure}
\includegraphics[width=1.0\linewidth,clip=true]{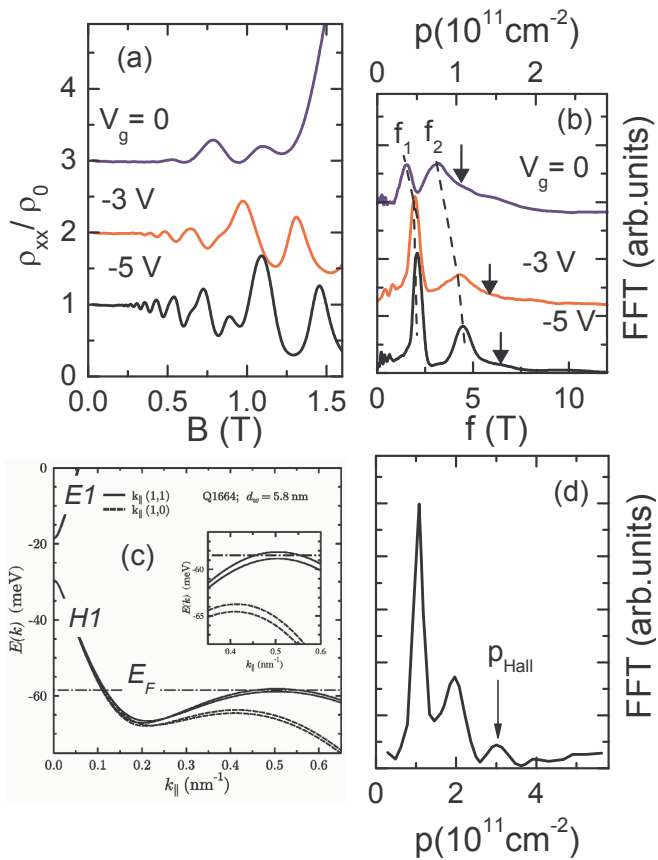}
\caption{The SdH oscillations for some gate voltages (a)
and their Fourier spectra  (b).  The calculated energy spectrum
(c) and the Fourier spectrum of the SdH oscillations (d)
from Ref.~\onlinecite{Ortner02}. Arrows in (b) correspond to $p_H$.
}\label{f3}
\end{figure}

Another way to determine the hole density is the analysis of the  SdH
oscillations. The experimental  SdH oscillations of $\rho_{xx}$ are
shown for several gate voltages in Fig.~\ref{f3}(a), while the
corresponding Fourier spectra are presented in Fig.~\ref{f3}(b).	Two
maxima  with frequencies $f_1$ and $f_2$, which are shifted with the
gate voltage can be easily detected in the Fourier
spectra.\footnote{Note that in order to find the parameters of the
energy spectrum from the SdH oscillations one should make the Fourier
analysis at low enough magnetic field when the oscillations of the
Fermi energy with the magnetic field can be neglected, i.e., when
$\delta \rho_{xx}/ \rho_{xx}\ll 1$.} Noteworthy is that the ratio of
the frequencies is close to $2$ (see the inset in Fig.~\ref{f2}). So,
the Fourier spectra are analogous to the case when the spin splitting
of the Landau levels manifests itself with the magnetic field increase.
In such a situation the carrier density should be determined as
$p_{\,\text{SdH}}= f_1\times (e/\pi\hbar)$. If this is true in our
case, we obtain $p_{\,\text{SdH}}$ which is significantly less than the
hole density $p_{\,\text{H}}$ obtained from the Hall effect. For
example, inspection of Fig.~\ref{f3}(b) reveals $f_1\simeq 2$~T for
$V_g=-3$~V that yields $p_{\,\text{SdH}}\simeq 0.95\times
10^{11}$~cm$^{-2}$, whereas the Hall effect for this gate voltage gives
much larger value, $p_{\,\text{H}}\simeq 1.4\times10^{11}$~cm$^{-2}$
(see Fig.~\ref{f2}).

Before discussing this discrepancy, let us point out the fact  that
very similar results were obtained in Ref.~\onlinecite{Ortner02} for
the HgTe quantum well of the same nominal width but with somewhat
larger Hall density, $p_{\,\text{H}}=3\times 10^{11}$~cm$^{-2}$ [see
Fig.~\ref{f3}(d)]. As seen two maxima in the Fourier spectrum with the
ratio of the frequencies of about two were observed also in that paper.	
Therewith the hole density found from the SdH oscillations turns out to
be less than the Hall density.

There are several possibilities to get out of this disparity,
$p_{\,\text{H}}>p_{\,\text{SdH}}$. The authors of
Ref.~\onlinecite{Ortner02} attributed this discrepancy to the
peculiarity of the valence band energy spectrum. The energy spectrum
calculated for the actual energy range in Ref.~\onlinecite{Ortner02} is
presented in Fig.~\ref{f3}(c). As seen  there are additional maxima on
the dispersion of the valence band $H1$ in the (1,1) direction at
$k_\parallel\simeq 0.5$~nm$^{-1}$. These maxima are situated $28$~meV
lower than the top of  the main maximum  at $k_\parallel=0$. In
accordance with the calculation, the authors of
Ref.~\onlinecite{Ortner02} reasoned that the experimental peak
corresponding to $p=1\times 10^{11}$~cm$^{-2}$ is two merged peaks
centered at $p=1.01\times 10^{11}$~cm$^{-2}$ and $0.97\times
10^{11}$~cm$^{-2}$, which originate from the spin-orbit split $H1-$ and
$H1+$ subbands at $k_\parallel\approx 0$. The peak at $p=1.95\times
10^{11}$~cm$^{-2}$ corresponds to the sum of the hole densities in
these subbands. So, the hole density $p_{\,\text{SdH}}$ found from the
SdH oscillations is about $2\times 10^{11}$~cm$^{-2}$ according to the
interpretation given in Ref.~\onlinecite{Ortner02}. The difference
between the  Hall density, $p_\text{Hall}=3\times 10^{11}$~cm$^{-2}$
[see Fig.~\ref{f3}(d)] and SdH density, $p_{\,\text{SdH}}=2\times
10^{11}$~cm$^{-2}$, which is about of $1\times 10^{11}$~cm$^{-2}$, has
been attributed to the holes in the four secondary maxima at
$k_\parallel=0.5$~nm$^{-1}$. According to the authors:\cite{Ortner02}
``A peak in the Fourier spectrum due to these holes could be expected;
however, this should occur at a very low frequency corresponding to
$p=0.24\times 10^{10}$~cm$^{-1}$ and is therefore not observed''. It is
clear that in the framework of this model the decrease of the total
hole density down to $2\times 10^{11}$~cm$^{-2}$ should lead to the
disappearance of the contribution of the carriers from the secondary
maxima and, thus, should lead to agreement between the Hall and SdH
densities. Unfortunately, the structure studied in
Ref.~\onlinecite{Ortner02} was ungated therefore the authors could not
control the density of the carriers to check such interpretation. So,
the results of Ref.~\onlinecite{Ortner02} does not seem very
conclusive.

In the structures investigated in the present paper, the difference
between the density  found from the SdH oscillations within proposed
model and the density found from the Hall effect is observed over the
whole gate voltage range, where the hole density changes from
$0.6\times 10^{11}$~cm$^{-2}$ to $1.7\times 10^{11}$~cm$^{-2}$. Thus,
the described above interpretation does not correspond to our case.

The only model that describes our results is as follows. As in
Ref.~\onlinecite{Ortner02}, we suppose that the subband of spatial
quantization $H1$ is split by spin-orbit interaction into two subbands,
$H1+$ and $H1-$, due to asymmetry of the quantum well.  But in contrast
to Ref.~\onlinecite{Ortner02}, we will believe that the spin-orbit
splitting is so large that the two maxima evident in our Fourier
spectra originate just from these $H1+$ and $H1-$ subbands. Under this
assumption the hole densities in the split subbands should be found as
$p_{1,2}=f_{1,2}\times(e/2\pi\hbar)$, where the indexes $1$ and $2$
correspond to the $H1+$ and $H1-$, respectively. The factor two in the
denominator is due to the absence of ``spin'' degeneracy of the split
subbands. The total density in this case is the sum of $p_1$ and $p_2$;
$p_{tot}=p_1+p_2$. The results of such a data treatment  are presented
in Fig.~\ref{f2} within the gate voltage range from $-8.5$~V to
$+0.5$~V, where we were able to determine reliably the frequencies for
both peaks in the Fourier spectrum. One can see that the values of
$p_{tot}$ coincide with $p_{\,\text{H}}$, to within experimental error.
At $V_g>0.5$~V, we could determine the frequency of only low-frequency
Fourier component resulting from the quantization  of $H1+$ subband. In
this case the hole densities in the second subband, $H1-$, was found as
$p_2=p_{\,\text{H}}-p_1$  and is shown  in Fig.~\ref{f2} by the open
circles. As seen these data match the data obtained directly from the
Fourier spectra at $V_g<0.5$~V well.

The inset in Fig.~\ref{f2} shows that the ratio of the hole densities
in spin-orbit split subbands is practically independent of the gate
voltage and consists of about $2$ over the whole gate voltage range. So
large ratio is an evidence of the giant spin-orbit splitting.\footnote{
It might be expected that the existence of two types of the carries
corresponding to the two spin-orbit split subbands should reveal itself
in the magnetic field dependences of Hall coefficient and longitudinal
resistance. However, it is easy to check that for such a ratio between
the densities and with the ratio between the mobilities of about
$1.5-2$ times, the change of $R_{\,\text{H}}$ and $\rho_{xx}$ in
classical magnetic fields should be about $(4-7)$ \% only. Really, the
changes of $R_{\,\text{H}}$ and $\rho_{xx}$ with magnetic field about
$(3-5)$\% are observed, however it is impossible to find four
parameters unambiguously from these dependencies.}

\begin{figure}
\includegraphics[width=0.96\linewidth,clip=true]{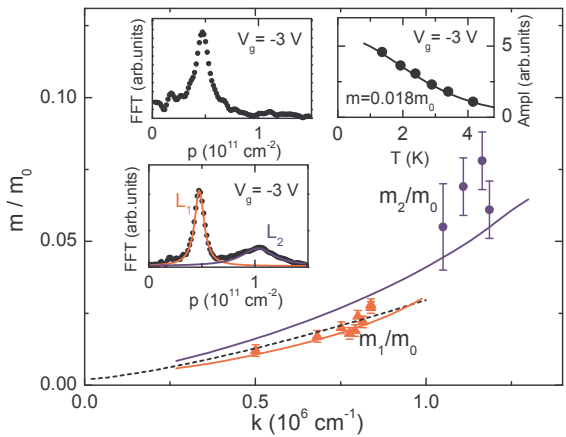}
\caption{ The values of the hole effective mass in the subbands $H1+$ (triangles)
and $H1-$ (circles) found at different quasimomentum $k_{1,2}=\sqrt{p_{1,2}/4\pi}$. The dotted line is the interpolation
dependence $m_1(k)$. The solid lines are the result of the
theoretical calculation (see text). The left upper   and lower insets show the
Fourier spectrum of the oscillations of $\rho_{xx}$ taken at $V_g=-3$~V
within the magnetic field ranges $(0.1-0.4)$~T and $(0.1-1.1)$~T, respectively.
The lines in the lower inset are the best fit by the two Lorentzians.
The right inset is the temperature dependence of the oscillations
amplitude at $B=0.3$~T (symbols) and the result of the best
fit by the Lifshits-Kosevich formula\cite{LifKos55} with $m=0.018m_0$ (line). }\label{f4}
\end{figure}

At first sight the large spin-orbit splitting of the energy spectrum in
the nominally almost symmetrical structure [see the inset in
Fig.~\ref{f1}(a)] seems very surprising. However, one must take into
account the peculiarity of MBE growth process of the
Hg$_{1-x}$Cd$_{x}$Te/HgTe heterostructures. There is an overpressure of
Te during the growth process, therefore the Hg vacancies  are presented
in the structure at the end of growth. They are the acceptors in
Hg$_{1-x}$Cd$_{x}$Te, therefore  major carriers in the quantum well
should be the holes. Nevertheless,  the heterostructures demonstrate
$n$-type conductivity after evacuation from the growth chamber as a
rule. It is believed that this is because the mercury overpressure
remains in the chamber after completion of growth. During the cooling,
the vacancies of the mercury are annealed. When the cooling time is
small, the upper barrier is converted to the $n$-type, while the lower
barrier can remain of $p$-type. In this case, the quantum well is
brought into the $p$-$n$ junction and the strong electric field of the
junction should lead to spin-orbit splitting due to the Rashba
effect.\cite{Rash84}

Thus, the analysis of the gate voltage dependences of the Hall density
and SdH oscillations shows that we are dealing with the structures
whose valence band is strongly split due to spin-orbit interaction.
Knowing this we are in position to study the spectrum in more detail.

In order to do this we have determined the hole effective mass
analyzing the temperature dependence of the amplitude of the SdH
oscillations. As seen from the left upper inset in Fig.~\ref{f4}, the
main contribution to the SdH oscillations at low enough magnetic field
comes from the one spin-orbit split subband $H1+$, which gives the
lower frequency to the Fourier spectrum. Thus, the effective mass $m_1$
found from  the SdH oscillations within this magnetic field range will
correspond to the mass in the $H1+$ subband. As an example, the
temperature dependence of the amplitude of SdH oscillations at
$B=0.3$~T measured at $V_g=-3$ V is shown in the right inset in
Fig.~\ref{f4}. The fit of this dependence by the Lifshits-Kosevich
formula\cite{LifKos55} (shown by the  line) gives $m_1= (0.018\pm
0.003)m_0$. Such analysis performed for different gate voltages gives
the quasimomentum dependence of the effective mass $m_1$ which is
plotted in Fig.~\ref{f4} by the triangles. One can see that $m_1$
increases significantly with $k$ increasing.

\begin{figure}
\includegraphics[width=\linewidth,clip=true]{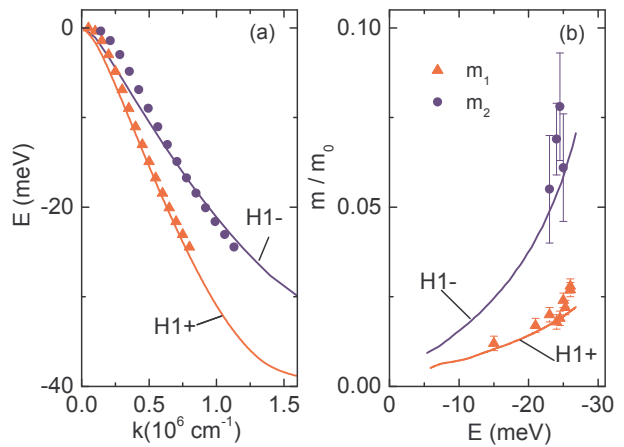}
\caption{The energy spectrum (a) and the effective mass plotted as a function of the energy (b)
for the $H1+$ and $H1-$ hole subbands. Symbols are  restored from the experimental data, the solid lines
are the results of theoretical calculation with the
taking into account the electric field in the well.
}\label{f5}
\end{figure}

The experimental determination  of the effective mass in the second
subband, $H1-$, is much more difficult. With this aim, we have
decomposed the Fourier spectra for every temperature by fitting them by
two Lorentzians, $L_1$ and $L_2$ (see the left lower inset in
Fig.~\ref{f4}). Then, after the inverse Fourier transformation of $L_2$
we obtained the oscillations coming from the $H1-$ subband. From the
temperature dependence of the amplitude of these oscillations we have
found the effective mass $m_2$. The results of such a data treatment
are shown in Fig.~\ref{f4} by the circles.

Knowing $m$~vs~$k$ dependence and assuming an isotropic energy spectrum
one can restore the dispersion $E(k)$: $E(k)=\int_0^k k/m(k) dk$.
Because $m_1$ is measured within a wider interval of $k$, we have
firstly obtained $E(k)$ for the $H1+$  subband. Using the interpolation
dependence $m(k)$  shown in Fig.~\ref{f4}  by the dotted curve we have
obtained the dependence $E(k)$, which is depicted in Fig.~\ref{f5}(a)
by the triangles. To restore the dispersion of the $H1-$ subband we use
the fact that the ratio of the hole densities in the $H1-$ and $H1+$
subbands is about $2$ over the whole gate voltage range [see the inset
in Fig.~\ref{f2}]. So the dispersion of the $H1-$ subband has been
obtained from the $H1+$ dispersion by scaling with the factor
$\sqrt{2}$ in the $k$ direction. The value of spin-orbit splitting is
really gigantic: for example, it is about $8$~meV at the Fermi energy
$20$~meV, i.e., approximately $40$~\%, Now, when we have restored the
$E$~vs~$k$ dependence we can find the energy dependence of the
effective masses [see Fig.~\ref{f5}(b)]. It is seen that $m_1$
increases strongly with the energy and $m_2$ is $2-3$ times larger than
$m_1$.

To compare the experimental results with the theory, we have calculated
the energy spectrum within framework of the six-band \emph{kP} model
taking into account the lattice mismatch between the
Hg$_{1-x}$Cd$_{x}$Te layers forming the quantum well and CdTe buffer
layer.  The calculations have been performed within framework of
isotropic approximation using the direct integration technique as
described in Ref.~\onlinecite{Lar97}. The run of the electrostatic
potential has been obtained self-consistently from the simultaneous
solution of the the Schr\"{o}dinger and  Poisson equations.  The donor
and acceptor densities in the upper and lower barriers, respectively,
were supposed to be equal to $3\times 10^{17}$~cm$^{-3}$. The other
parameters were the same as in Refs.~\onlinecite{Zhang01,Novik05}.

\begin{figure}
\includegraphics[width=0.8\linewidth,clip=true]{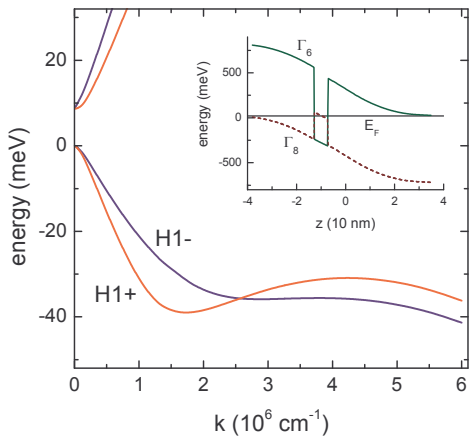}
\caption{The results of self-consistent calculation of the energy spectrum of
the conduction and valence bands. The inset shows the energy diagram of
the structure on the assumption  that acceptor and donor densities in
the lower and upper barriers, respectively, are $3\times
10^{17}$~cm$^{-3}$.
}\label{f6}
\end{figure}

The calculated energy diagram  and the dispersion law $E(k)$ for the
heterostructure H724 are shown within a wide $k$ range in
Fig.~\ref{f6}. As clearly seen the energy spectrum of valence and
conduction bands is strongly split due to the Rashba effect, therewith
the positions of the branches $H1+$ and $H1-$ are interchanged at $k
\simeq2.5\times 10^6$~cm$^{-1}$.

To compare theoretical and experimental results, we have depicted the
calculated dependences $m(k)$, $E(k)$ and $m(E)$ on the same figures,
where the experimental data are presented [see Fig.~\ref{f4},
Figs.~\ref{f5}(a) and \ref{f5}(b), respectively]. It is evident that in
the range, where the experimental data were obtained,  they are in good
agreement with the theoretical curves suggesting the adequacy of the
model used.

\begin{figure}
\includegraphics[width=\linewidth,clip=true]{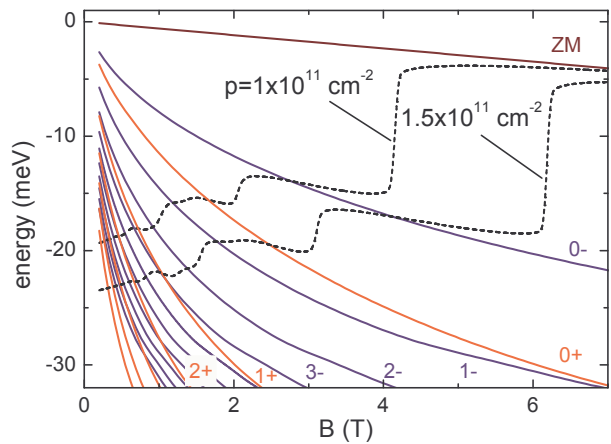}
\caption{(Color online) The  Landau levels calculated within
semi-classical approximation as a function of  magnetic field. The dotted lines represent
the Fermi level for two hole densities calculated with Landau levels broadening of $1$~meV.
The Landau levels are labeled with quantum numbers $N$, and the signs $+$ and $-$ indicate
from which subband, $H1+$ or $H1-$, the levels are originated. The zero-mode Landau level is labeled as ZM.
}\label{f7}
\end{figure}

Let us now turn to the peculiarity of the quantum Hall effect mentioned
in the beginning of this section and evident as the absence of the
quantum Hall plateaus with the numbers $3$ and $6$ [see
Fig.~\ref{f1}(b)]. Interestingly, the absence of the same plateaus was
also observed but not commented and discussed in
Ref.~\onlinecite{Ortner02} (see Fig.~7 in that paper). To understand
this feature, one should  calculate the Landau levels. It can be in
principle done with the use of one of technique presented in the
literature.\cite{Novik05,Zholudev12} However, the qualitative
explanation can be already obtained within much simpler semi-classical
quantization approximation. The energy of the $N$-th Landau level $E_N$
can be obtained in this case by substitution of $(2N+1)/l^2$ instead of
$k^2$ in the dependence $E(k)$, where $l=\sqrt{\hbar/eB}$ is the
magnetic length. Moreover, it must be taken into account that there is
an additional zero-mode (ZM) Landau level  for such a spectrum, whose
position is almost independent of the magnetic
field.\cite{Novik05,Buttner11,Zholudev12,minkov13-2} In Fig.~\ref{f7},
we have depicted the fan-chart diagram calculated in such a manner for
the spectrum shown in Fig.~\ref{f5}(a). The dotted lines in this figure
are the Fermi level calculated with the Landau levels broadening of
$1$~meV for two hole densities, $p=10^{11}$~cm$^{-2}$ and $1.5\times
10^{11}$~cm$^{-2}$. One can see that the energy distances between the
levels $1-$ and $0+$, and the levels $3-$ and $2+$ near the Fermi level
within this hole density range are noticeable less than that between
the other levels. Therefore, the localized states between these Landau
levels can be absent so the $\rho_{xy}$ plateaus with the numbers $3$
and $6$ have not to be observed.

Another feature mentioned above is that the $p$~vs~$V_g$ dependence is
flattened at $V_g\simeq -3.5$~V when the density reaches the value
$1.5\times 10^{11}$~cm$^{-2}$  with the lowering gate voltage
(Fig.~\ref{f2}). There are two possibilities to explain this behavior.
The first one results from the peculiarity of the energy spectrum. It
is easy to estimate from Fig.~\ref{f5}(a) that the Fermi level at
$p=1.5\times 10^{11}$~cm$^{-2}$ lies at the energy of about $-25$~meV.
As seen from Fig.~\ref{f6}, this value is close to the energy of
secondary maxima in the dependence $E(k)$ located at $k\simeq 4\times
10^6$~cm$^{-1}$. Because of the large effective mass in these maxima
the sinking speed of the Fermi level decreases strongly when these
states are being occupied at $V_g<-3.5$~V. In the presence of potential
fluctuations, these states can be localized and, hence, they will not
contribute to the conductivity. In this case, the Hall density will
correspond to the hole density in the main maximum (while the total
charge of carriers in the well will be determined by the density in all
the maxima). In favour of this conclusion is the fact that the
flattening of the dependence $p(V_g)$ is observed in all three
structures under study.

The second possibility to explain the feature under discussion is
existence of localized states in the lower barrier which start to be
occupied with the  decreasing gate voltage leading to the same effect
in the dependence $p(V_g)$ just at $V_g\simeq -3.5$~V. We cannot
exclude this mechanism at the moment.

\section{Conclusion}

\label{sec:concl}

We have studied the transport phenomena in the gated quantum well
Hg$_{1-x}$Cd$_x$Te/HgTe of $p$-type conductivity with the normal energy
spectrum. Analyzing the data we have reconstructed the dispersion law
near the top of the valence band at $k\lesssim 10^6$~cm$^{-1}$. It has
been shown that the hole energy spectrum is strongly split by the
spin-orbit interaction, so that the ratio of the holes in the split
subbands is approximately equal to two. It has been shown that the
energy spectrum is strongly non-parabolic, i.e., the hole effective
masses significantly increase with the energy increase. These results
are well described in the framework of the \emph{kP} model if one
supposes that the lower barrier remains of $p$-type, while the upper
one is converted to the $n$-type after the growth stop so that the
quantum well is located in a strong electric field of $p$-$n$ junction.

Noteworthy is that the above conclusion on the adequacy of the
\emph{kP} model for description of the hole energy spectrum is opposite
to the conclusion made in our previous paper\cite{Minkov13} on the wide
quantum wells with the inverted spectrum. It was there shown that  the
valence band spectrum is electron-like near $k=0$ so the top of the
band is located at $k\neq 0$. The key result of the paper, however, was
that the experimental and calculated hole spectra being in qualitative
agreement were strongly different quantitatively within whole range of
experimentally accessible quasimomentum values, $k\lesssim 1.7\times
10^6$~cm$^{-1}$.

One of the possible reasons why the theory works well in the narrow
HgTe quantum wells and does not explain the data in  wide wells is the
following. The top of the valence band is formed from the different
subbands of spatial quantization in narrow and wide quantum wells. The
top is the $H1$ subband in the first case and  the $H2$ subband in the
second one. Thus it turns out that the standard \emph{kP} model
describing the dispersion of the $H1$ subband  fails to describe the
spectrum of $H2$ subband for whatever reason. The other reason concerns
the band gap. In the quantum wells investigated in the present paper,
the valence and conduction bands are separated by the gap. The systems
investigated in Ref.~\onlinecite{Minkov13} are semimetallic, therefore
electrons and holes can coexist in such a case. A workability of
single-particle approximation in such a situation can be really
questionable and a many-particle approach could be more adequate.

\section*{Acknowledgments}

We thank A.~Ya.~Aleshkin and M.~S.~Zholudev for helpful discussions.
Partial financial support from the RFBR (Grant Nos. 12-02-00098 and
13-02-00322) are gratefully acknowledged.

\section*{References}
%

\end{document}